\documentclass[preprint,,showpacs,aps]{revtex4}

\usepackage{graphicx}
\usepackage{dcolumn}
\usepackage{bm}

\begin{document}

\preprint{Preprint submitted to Appl. Phys. Lett.\space\today}

\title{Ultratransparent glass-ceramics: the structure factor and the quenching of the Rayleigh scattering.}

\author{M. Mattarelli$^{a)}$,  M. Montagna\footnotetext {$^{a)}$Corresponding author. Fax: +39-0461-
    881696;  e-mail: mattarel@science.unitn.it}}
\affiliation{Dipartimento di Fisica, CSMFO group,
Universit$\acute{a}$ di Trento, Via Sommarive 14, I-38100 Trento,
Italy.}
\author{P. Verrocchio}
\affiliation{Dipartimento di Fisica and SOFT, Universit$\acute{a}$
di Trento, Via Sommarive 14, I-38100 Trento, Italy.}

\date{\today}

\begin{abstract}
Glass-ceramics with nanocrystals  present a transparency higher
than that expected from the theory of Rayleigh scattering. This
ultra-transparency is attributed to the spatial correlation of the
nanoparticles. The structure factor is calculated for a simple
model system, the random sequential addition of equal spheres, at
different volume filling factor. The spatial correlation given by
the constraint that particles cannot superimpose produces a
diffraction peak with a low $S(q)$ in its low-$q$ tail, which is
relevant for light scattering. The physical mechanism producing
high transparency in glass-ceramics is demonstrated to be the low
density fluctuation in the number of scatterers.

\end{abstract}

\pacs{78.20 -Bh; 78.35.+c; 81.05.Pj}

\keywords{}

\def\@oddhead{\REVTeX{}4\hfill\space Preprint submitted to Phys.
Rev. Lett.\space\today}
\makeatother

\maketitle

Glass-ceramics with nanometric size of the crystals are of great
interest for photonics, since they combine the spectroscopic
properties of crystals with the mechanical properties of glasses
\cite{Wang,tick,victor,almeida,mortier}. Many glass-ceramics, in
particular those obtained from oxyfluoride glasses, have shown a
transparency similar to that of glasses, much higher than that
expected from the theory of Rayleigh scattering
\cite{tick,victor}. Recently, Edgar et al.  measured the optical
extinction coefficient for fluorozirconate glasses containing a
volume fraction of about 0.2 of barium chloride nanoparticles,
finding that the glass ceramics are about a factor of six more
transparent than predicted by the Rayleigh scattering theory,
based on estimates of particle sizes from X-ray diffraction and
electron microscopy measurements \cite{edgar}. They modelled the
electric response of the material within a discrete dipole
approximation, which accounts for the particle-particle
interaction. However, they did not succeed in reproducing the
observed quenching of the attenuation. In our opinion the
dipole-induced-dipole effects are not crucial for the scattering
intensity. The particles, in a first approximation, can be
considered independent and excited by a homogeneous external
field. The main effect is related to their spatial arrangement and
the consequent interference pattern of the scattered field. Tick
suggested that the low attenuation is related to this effect, but
quantitative calculation are lacking. Hopper treated the case of a
spinodal decomposition and Hendy found a $q^{8} \sim\lambda^{-8}$
dependence of the Rayleigh cross section in the low $q$ limit, $q$
being the wavevector of the light ($q=2\pi/\lambda)$ \cite{hendy}.

In the absence of any spatial correlation the scattering
attenuation is given by
\begin{equation}
\alpha= \sum_{i}\sigma_{i},
\end{equation}
where the sum in on all the particles in the volume unit and
$\sigma_{i}$ is the cross section of Mie scattering of the i-th
particle. If the interaction among particles is neglected, i.e.
each particle is considered as surrounded by a homogeneous medium,
and if all particles are equal, Eq (1) reduces to $\alpha=\rho_N
\sigma$, where $\rho_N$ is the particle density. Furthermore, if
the size of the particle is much smaller than the wavelength of
the light, Rayleigh scattering occurs. The cross section for
Rayleigh scattering by a not absorbing spherical particle of
radius $R$ and refractive index $n_{p},$ embedded in a not
absorbing medium of refractive index $n_{m}$, is given by
\cite{vandehulst}:
\begin{equation}
\sigma _{Ray}=\frac{8}{3}\pi k^{4}R^{6}\left(
\frac{m^{2}-1}{m^{2}+2}\right) ,
\end{equation}
where $m=n_{p}/n_{m}$, $k=2\pi /\lambda =2\pi n_{m}/\lambda _{0}$
is the wave vector of the light of wavelength $\lambda $ in the
medium and $\lambda _{0}$ in vacuum.

In the presence of spatial correlation, interference effects occur
and the total scattered field at any angle ($\theta$) of
scattering, or at any exchanged wavevector
$q=(4\pi/\lambda)\sin(\theta/2)$, is the sum of all individual
fields. For $N$ equal particles excited by a plane wave, the
individual scattered fields are given by:
\begin{equation}
\textbf{E}_{i}(\textbf{q})=\textbf{E}_{0}(\textbf{q})\exp(i\textbf{qr}_i),
\end{equation}
and the scattered intensity is given by:
\begin{equation}
\textbf{E}(\textbf{q})^2 = N \textbf{E}_{0}(\textbf{q})^2 S(q),
\end{equation}
where $S(q)$ is the structure factor of the system:
\begin{equation}
S(q)=\mid \sum_{i}\exp(i\textbf{qr}_i)\mid ^2/N.
\end{equation}

We will show that any spatial correlation lowers significantly
$S(q)$ at the small $q$-values with respect to the value $S(q)=1$
of the uncorrelated system. The attenuation of the light will
depend on the scattering at all angles ($0<\theta<\pi$) or
$q$-values ($0<q<4\pi/\lambda$).
 Furthermore, we will show that for small $q$-values,  i.e.
the ones which are relevant to Rayleigh scattering, the structure
factor is nearly constant, i.e. $S(q)\simeq S(q=0) \equiv S_0$.
Therefore the attenuation, in the presence of a particle
correlation measured by $S(q)$ is well approximated by $\alpha_c =
\alpha_S S_0=\rho_N \sigma_{Ray} S_0$.

An unavoidable correlation among equal spherical  hard particles
 is given by the impossibility of interpenetration.
Interestingly, this effect may enforce non-trivial spatial
  correlations even on length-scales larger than $2R$. A simple process
  yielding three dimensional systems subjected only to such hard-sphere constraint
is given by the random sequential addition (RSA) \cite{torquato}.
In such process, the $n$-th particle is placed randomly within a
cubic box with periodic boundary condition only if it finds the
space left free from the previous $n-1$ particles.  The
  correlation among the particles increases when the filling factor
  $\phi=4/3\pi R^3 \rho_N $, where $R$ is the particle radius, grows.
For $\phi=0$, there is no spatial correlation, while the maximum
filling factor for RSA in 3-D is $\phi_{sat}\simeq0.3828$
\cite{torquato}.

Figure 1 shows the calculated $S(q)$ for various values of $\phi$,
obtained by averaging over $30$ realizations of the RSA process
with $N=10000$ particles. As $\phi$ increases, a diffraction peak,
located at $q_{p}\simeq 2\pi/2R$ becomes more and more well
defined. The radial distribution function $g(r)$ (not shown here)
is zero for $r<2R$ and presents a single peak at $r_p \gtrsim 2R$.
The radial correlation decays very fast
 with $r$ for $r>2R$ for any $\phi$, similarly
to the case near $\phi_{sat}$ ($\phi=0.3812$), described by
Torquato et al. \cite{torquato}.

\emph{Figure 1 around here }

The quenching of the $S(q)$ in the low-$q$ region is shown in more
detail in the inset of Fig. 1. When plotted as a function of $q^2$,
the parabolic behavior \cite{torquato}
\begin{equation}
S(q)=S_0+ aq^2
\end{equation}
 appears as a
straight line. Here $S_0$ is the low-$q$ limit, being
$q_{min}=2\pi /L$, where $L$ is the edge of the cubic box, the
minimum $q$-value accessible in the calculations. The fitted $S_0$
values are reported in Fig. 2, together with the value found near
the saturation density \cite{torquato}.

 A quantitative prediction can be obtained by studying the
  simplest form for $g(r)$ which takes into account both the
  hard-sphere constraint and some possible correlation over a
  lenght-scale $\xi$,
\begin{eqnarray}
g(r) &=& 0 \quad r < 2R \nonumber \\
g(r) &=& 1 + A e^{-r/\xi} \quad r>2R.
\end{eqnarray}
In fact with that choice one finds:
\begin{equation}
S_0 = 1 - 8 \phi,
\end{equation}
which suggests that the increase of the filling factor is the main
factor which determines the drop of {\bf $S_0$}.
Therefore, the very short range correlation of our model system,
which only avoids particle superposition, is able to reduce the
scattering intensity by about an order of magnitude for $\phi$ in
the range of $0.2-0.3$, values that are typical of glass-ceramics.
Furthermore, no medium or long range correlation seems to be
needed for quenching Rayleigh scattering. But, in fact, some
degree of long range order is produced. A deeper insight is
obtained by the following study. The box containing $N$ particles
($N$=10000 in our case) can be decomposed in $n$ cubic smaller
boxes, each containing $M=N/n$ particles, on average. In the
absence of any particle correlation, a Poisson distribution in the
number $M_j$ of particles will be present. For large values of
$M$, it is well approximated by a Gaussian distribution with
variance $\sigma_M^2 = M$. In the inset of Fig. 2 we show the
results for $n=125$, $M=80$ and for two values of the volume
filling factors, $\phi=0.05$ and $\phi=0.3$. The variance of the
particle number is smaller than $M$ and decreases as the filling
volume increases. This can be easily understood, since in the RSA
method of box filling the $n$-th particle will have larger
probability of being accommodated in a box with smaller particle
density. This will produce an equalization of density, reducing
its fluctuations  over large scales to lower values than those
that would be present for a random distribution. Figure 2
summarizes the results of the calculations for different
$n$-values of the box partition, $n=125$, 64, 27 and 8, showing
the normalized variance ${\sigma'_M}^2={\sigma_M}^2/M$.

\emph{Figure 2 around here }

The above feature is not limited to RSA systems. As a matter of
  fact, in a system of classical particles, $S_0$ is always related to
  the fluctuations of the number of particles $M$ within spheres of
  (large) radius $l$ by the relation \cite{hansen}:
\begin{eqnarray}
\frac{\langle M^2 \rangle - \langle M \rangle^2}{\langle M
\rangle}
&=& 1 + 4 \pi \rho_N \int_0^l dr \,r^2 \left(g(r)-1 \right) \nonumber \\
&\to  & S_0 \, ,
\end{eqnarray}
where the limiting value is approached when $l \to \infty$ (this
explains the observed size dependence of $\sigma'_M$ on the
sampling size M in Fig. 2).

In real systems, as also suggested by the model of Eq. 7, $S_0$
can be much lower than what is found in RSA \cite{tick,hendy}. A
naive argument, which basically exploits Eq. 9, allows accounting
for such a decrease. In fact, even though the processes of
nucleation and coarsening are very system dependent, the growth of
nanosized particles usually occurs by diffusion within limited
spatial region, each particle having its own basin for gathering
material. In a volume with a characteristic distance of the order
of $1 \mu m$ relevant for Rayleigh scattering, we will find a
certain number $N_{BB}$ of the building blocks (BB) that will
compose the nanoparticles.
If we assume that $M=N_{BB}/n_{BB}$ identical nanoparticles, each
one containing $n_{BB}$ BB, are formed by limited diffusion within
the given volume, $M$ will have the same relative fluctuation as
$N_{BB}$: $\sigma_M/M=\sigma_{N_{BB}}/N_{BB}$. The fluctuations on
the particle number will therefore be proportional to $1/n_{BB}$
and, by Eq. 9, so will be $S_0$ and the intensity of the scattered
light. Note that since the cross section $\sigma_{Ray}$ is
proportional to the square of the particle volume (see the $R^6$
dependence in Eq. 2), $\sigma_{Ray} \rho_N$ is proportional to the
particle volume (assuming that all the BB have been precipitated
in particles and $n_{BB}M=N_{BB}$), and finally $\alpha_c
=\sigma_{Ray} \rho_N S_0$ becomes size independent. Therefore, no
increase of attenuation is expected with the increase of the
particle size, since the increase of the cross section for
Rayleigh scattering is exactly compensated by the decrease of the
structure factor.

 This simple model does not account for other
important effects, as the size distribution of particles or the
formation of more ordered structures. In any case, the physical
mechanism producing high transparency in glass-ceramics even for
relatively high particle size is demonstrated to be the low
fluctuation in the number of scatterers, due to spatially limited
diffusion during coarsening.

The above results indicate two connected methods for estimating
the quenching of the Rayleigh scattering in a glass-ceramics. The
structure factor can be measured by Small Angle X-ray scattering.
If the measurements are extended, as far as possible, down to
low-$q$ side of the peak, the very low-$q$ S(q), $S_0$, should be
possibly obtained by extrapolation to the parabolic behavior of
Eq. 6. The fluctuations in the spatial particle distribution can
be measured by counting the particles using TEM tomography. The
decrease of Mie scattering is expected to be reduced by the factor
$S_0= {\sigma'_M}^2$ with respect to that calculated for a random
distribution of particles. The measure of light attenuation and of
the structure factor, combined with the calculation of the Mie
cross section for the single particle scattering, should
quantitatively confirm this explanation of the observed
ultra-transparency of glass-ceramics with nanocrystals.

The authors acknowledge M. Ferrari for useful discussions about
glass-ceramic materials.

\clearpage

\begin{figure}
\includegraphics [scale=1]{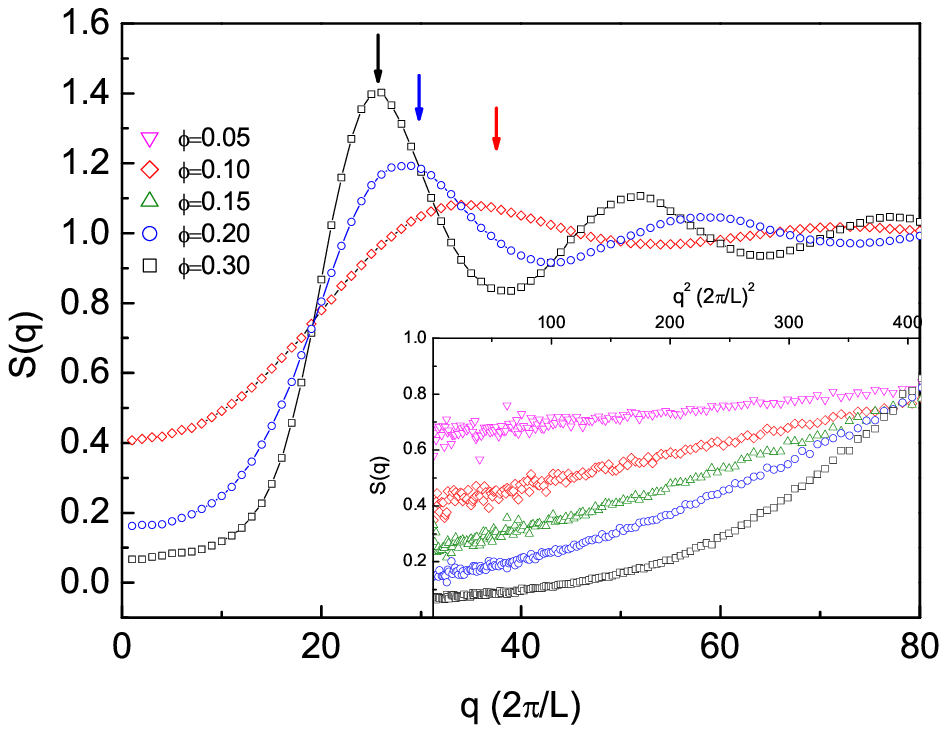}
\caption{Calculated structure factor for 3-D random sequential
addition of equal spherical particles and for different volume
filling ratios, $\phi$. The arrows indicate the values of
$q=2\pi/2R$, where $R$ is the radius of the spheres. The inset
shows the low-q part as a function of $q^2$.}\label{Figure 1}
\end{figure}

\begin{figure}
\includegraphics [scale=1]{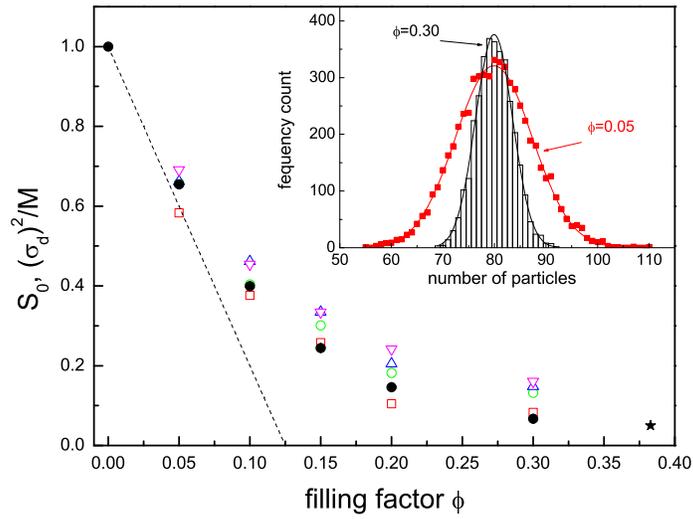}
\caption{Calculated low-q limit of $S(q)$ (dots). The star shows
the value ($S_0$ =.05) at the saturation density from ref.
\cite{torquato}. The dashed line shows the  $ \phi$-dependence
given by Eq. 8. Normalized variance of the particle density in
cubic boxes containing 10000/8 (open squares), 10000/27 (circles),
10000/64 (triangles) and 10000/125 (inverted triangles), on
average. Inset: distribution of number of particles in the 125
small cubes of size L/5 for  volume filling ratios $\phi$=0.05 and
0.3, calculated over 48 and 27 samples, respectively. The curves
are fits by Gaussian curves.}\label{Figure 2}
\end{figure}

\end{document}